# PROJECTIVE REPRESENTATIONS CONSTRUCTION FOR DIFFERENT POINTS OF BRILLOUIN ZONE. APPLICATION FOR SPACE SYMMETRY GROUPS P4$_1$2$_1$2 AND P4$_3$2$_1$2.

V.O. Gubanov[a], S.V. Koryakov[a]

[a] *Taras Shevchenko Kyiv National University, Department of Physics, Experimental Physics Division, 2 Akademika Hlushkov prosp., Kyiv, Ukraine, 03022 (e-mail: sergiy_koryakov@univ.kiev.ua)*

**ABSTRACT**

It was suggested method of constructing of irreducible projective representations in different points of Brillouin zone. Points Γ, Λ, Z, S, A, Σ, M, V, R, and X of space symmetry groups P4$_1$2$_1$2 and P4$_3$2$_1$2 were examined. At each of these points one- and two-valued irreducible projective representations of wave vector groups were constructed for two enantiomorphous modifications. Influence of time inversion at these points also was taken into consideration by means of Herring criterion.

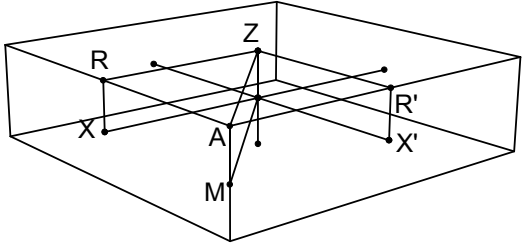

Fig. 1. Schematic drawing of Brillouin zone which corresponds to tetragonal crystals of space symmetry groups P4$_1$2$_1$2 and P4$_3$2$_1$2.

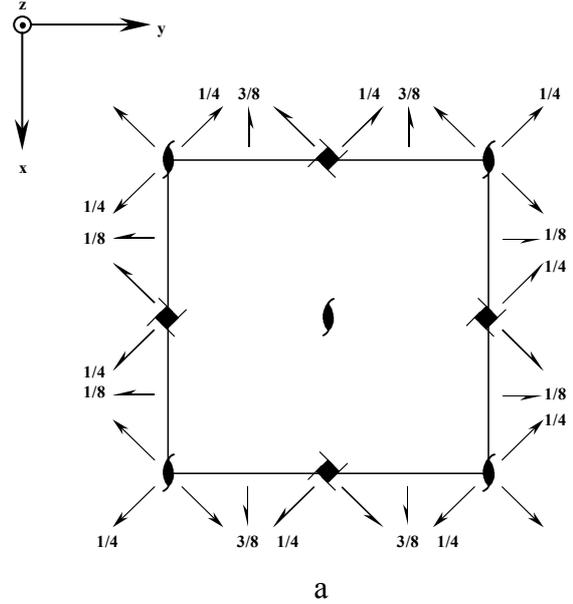

a

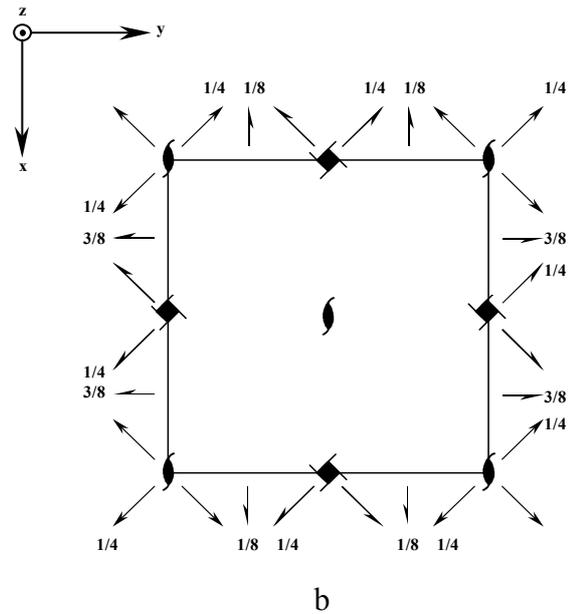

b

Fig. 2. Schematic drawing of space symmetry groups P4$_1$2$_1$2 (a) and P4$_3$2$_1$2 (b).

Space symmetry groups P4$_1$2$_1$2 and P4$_3$2$_1$2, are very interesting in practice, because they describe two enantiomorphous modifications of crystals, which characterised by appreciable anisotropy of the lattice and have quite high symmetry at the same time. So it is good opportunity to develop the general group theory method, that can allow to analyse energy states in such crystals.

In this paper we propose method of constructing of irreducible projective representations in different points of Brillouin zone of mentioned space groups. Here we examine points Γ, Λ, Z, S, A, Σ, M, V, R, and X, (Fig.1) but detailed calculation we will do only for points Γ, Λ, and Z, for other points procedure is similar, and we will represent final result (see appendix).

Let us construct irreducible projective representations $D_{\vec{k}}$ of wave vector groups $G_{\vec{k}}$. Representations $D_{\vec{k}}$ contain infinite number of components $D_{\vec{k}}(h)$ for elements $h \in G_{\vec{k}}$. Each of the elements can be written as $h = (\vec{\alpha} + \vec{a} \mid r)$. In this formula $r$ is rotation element, set of which creates point group $F_{\vec{k}}$, which is isomorfous to quotient system $G_{\vec{k}}$ by infinite invariant subgroup of translations, $\vec{\alpha}$ is a vector of non-trivial translation, which corresponds to the rotation element $r$, and $\vec{a}$ is a vector of trivial translation on one period of Brave lattice. $D_{\vec{k}}(h)$ can be defined as following:

$$D_{\vec{k}}(h) = e^{-i\vec{k}(\vec{\alpha} + \vec{a})} u(r) D(r), \qquad (1)$$

where $u(r) = u_{1\vec{k}}(r)$ in case when we consider states with integer spin, and $u(r) = u_{1\vec{k}}(r)u_2(r)$ for states with half-integer spin. Functions $u_{1\vec{k}}(r)$ reduce quotient system $\omega_1(r_2,r_1)$ to standard form $\omega'_1(r_2,r_1)$. Quotient system $\omega_1(r_2,r_1)$ is defined by properties of space group. Functions $u_2(r)$ reduce quotient system $\omega_2(r_2,r_1)$ to standard form $\omega'_2(r_2,r_1)$. Quotient system $\omega_2(r_2,r_1)$ determines transition into spin space. Product $u_{1\vec{k}}(r)u_2(r)$ reduces to standard form quotient system which determines transformation of spinors. $D(r)$ is the matrix of irreducible representation of projective class the same as corresponding quotient system.

For points Γ, Λ, and Z we will take the following wave vectors as a canonical: $\vec{k}_\Gamma = 0$, $\vec{k}_Z = -\dfrac{\vec{b}_1}{2}$, and $0 < \vec{k}_\Lambda < -\dfrac{\vec{b}_1}{2}$. So, when centre of Brilluin zone corresponds to point (0,0,0), we consider points with negative value.

Wave vector groups are equal for points Γ and Z, and are equal to full space group *G* with elements *g*. Point Λ will be examined separately, because wave vector group for this point is differ from *G* and is isomorphous to point group 4.

Primitive elements $h_i = g_i$ they determine wave vector groups, which can include only non-trivial translations, are defined as following (Fig.2): $h_1 = (0|e)$, $h_2 = \left(\dfrac{\vec{a}_1}{2}\bigg|c_2\right)$, $h_3 = \left(\dfrac{\vec{a}_1}{4}+\dfrac{\vec{a}_2}{2}+\dfrac{\vec{a}_3}{2}\bigg|c_4\right)$, $h_4 = \left(\dfrac{3\vec{a}_1}{4}+\dfrac{\vec{a}_2}{2}+\dfrac{\vec{a}_3}{2}\bigg|c_4^3\right)$, $h_5 = \left(\dfrac{3\vec{a}_1}{4}+\dfrac{\vec{a}_2}{2}+\dfrac{\vec{a}_3}{2}\bigg|(u_2)_1\right)$, $h_6 = \left(\dfrac{\vec{a}_1}{4}+\dfrac{\vec{a}_2}{2}+\dfrac{\vec{a}_3}{2}\bigg|(u_2)_2\right)$, $h_7 = \left(\dfrac{\vec{a}_1}{2}\bigg|(u'_2)_1\right)$, $h_8 = (0|(u'_2)_2)$ for group P4$_1$2$_1$2, and for group P4$_3$2$_1$2 - $h_1 = (0|e)$, $h_2 = \left(\dfrac{\vec{a}_1}{2}\bigg|c_2\right)$, $h_3 = \left(\dfrac{3\vec{a}_1}{4}+\dfrac{\vec{a}_2}{2}+\dfrac{\vec{a}_3}{2}\bigg|c_4\right)$, $h_4 = \left(\dfrac{\vec{a}_1}{4}+\dfrac{\vec{a}_2}{2}+\dfrac{\vec{a}_3}{2}\bigg|c_4^3\right)$, $h_5 = \left(\dfrac{\vec{a}_1}{4}+\dfrac{\vec{a}_2}{2}+\dfrac{\vec{a}_3}{2}\bigg|(u_2)_1\right)$, $h_6 = \left(\dfrac{3\vec{a}_1}{4}+\dfrac{\vec{a}_2}{2}+\dfrac{\vec{a}_3}{2}\bigg|(u_2)_2\right)$, $h_7 = \left(\dfrac{\vec{a}_1}{2}\bigg|(u'_2)_1\right)$, $h_8 = (0|(u'_2)_2)$. Here $\vec{a}_1$, $\vec{a}_2$, $\vec{a}_3$ are the main vectors of lattice, which are directed along axis *Oz*, *Ox* and *Oy* respectively. This choice of primitive elements is determined by standard choice of reference points for vectors of non-trivial translations in lattice.

Quotient systems $\omega_1(r_2,r_1)$ for points Γ and Z can be constructed with formula:

$$\omega_1(r_2,r_1) = e^{i(\vec{k}-r_2^{-1}\vec{k})\vec{a}_1}. \qquad (2)$$

It is obvious that all elements of $\omega_1(r_2,r_1)$ is equal to 1 for point Γ. It means that $\omega_1(r_2,r_1)$ is equal to standard quotient system of class $K_0$ $\omega'_{(0)}(r_2,r_1)$ for group 422.

**Table 1.** Quotient system of class $K_1$ for group P4$_1$2$_1$2 (point Z).

| $\omega_1(r_2, r_1)$ $r_2$ | $r_1$ | $e$ | $c_2$ | $c_4$ | $c_4^3$ | $(u_2)_1$ | $(u_2)_2$ | $(u'_2)_1$ | $(u'_2)_2$ |
|---|---|---|---|---|---|---|---|---|---|
| $e$ | | 1 | 1 | 1 | 1 | 1 | 1 | 1 | 1 |
| $c_2$ | | 1 | 1 | 1 | 1 | 1 | 1 | 1 | 1 |
| $c_4$ | | 1 | 1 | 1 | 1 | 1 | 1 | 1 | 1 |
| $c_4^3$ | | 1 | 1 | 1 | 1 | 1 | 1 | 1 | 1 |
| $(u_2)_1$ | | 1 | -1 | -i | i | i | -i | -1 | 1 |
| $(u_2)_2$ | | 1 | -1 | -i | i | i | -i | -1 | 1 |
| $(u'_2)_1$ | | 1 | -1 | -i | i | i | -i | -1 | 1 |
| $(u'_2)_2$ | | 1 | -1 | -i | i | i | -i | -1 | 1 |

**Note:** for group P4$_3$2$_1$2 quotient system $\omega_1(r_2,r_1)$ can be constructed in the same way.

These quotient systems belong to class $K_1$, because for commutating elements of this group the ratio $\dfrac{\omega(r_2,r_1)}{\omega(r_1,r_2)}$ equal to -1. Obtained quotient systems can be reduced to standard view by $u_{1\vec{k}}(r)$ (tab.2), which are calculated with formulas:

$$u_{1\vec{k}}(a^\kappa) = \dfrac{\omega_{an}^{\kappa/n}}{\omega_{a\kappa}}\varepsilon^\kappa, \qquad (3)$$

$$u_{1\bar{k}}(b^p) = -\frac{\omega^{p/2}(b,b)}{\omega_{am}\omega(a,a)}, \quad (4)$$

where $\omega_{an} = \omega(a,a)\omega(a,a^2)\ldots\omega(a,a^{n-1})$, $\kappa$ and $p$ – integer. We define $m$ from equation $\alpha' = e^{i\frac{2\pi m}{n}}$, where $\alpha' = \frac{\omega(b,a)\omega(a^{n-1},a)}{\omega_{an}^{2/n}\omega(a^{n-1},b)}$, $n$ is fold of main axis (in our case $n = 4$). For group P4$_1$2$_1$2 $m = 1$, and for group P4$_3$2$_1$2 $m = 3$.

$$\varepsilon = \frac{\omega(b,a^{n-1})}{\omega_{am}\omega(a,a)}(\alpha')^{1/2}.$$

Formulas (3), (4) is used for generating elements $b^p$ and $a^\kappa$. For elements $r = b^p a^\kappa$ following formula is used:

$$u_{1\bar{k}}(b^p a^\kappa) = \frac{u_{1\bar{k}}(a^\kappa)u_{1\bar{k}}(b^p)}{\omega_1(b^p,a^\kappa)}. \quad (5)$$

**Table 2** Functions $u_{1\bar{k}}(r)$

|  | Point | $e$ | $c_2$ | $c_4$ | $c_4^3$ | $(u_2)_1$ | $(u_2)_2$ | $(u'_2)_1$ | $(u'_2)_2$ |
|---|---|---|---|---|---|---|---|---|---|
| P4$_1$2$_1$2 | Z | 1 | $i$ | $-\varepsilon_8$ | $-\varepsilon_8^3$ | $-\varepsilon_8$ | $\varepsilon_8^3$ | -1 | $i$ |
| P4$_3$2$_1$2 | Z | 1 | $-i$ | $\varepsilon_8^3$ | $\varepsilon_8$ | $-\varepsilon_8^3$ | $\varepsilon_8$ | 1 | $i$ |

All quotient systems $\omega_1(r_2,r_1)$ are reduced to standard one $\omega'_1(r_2,r_1)$ with formula $\omega'(r_2,r_1) = \frac{\omega(r_2,r_1)u(r_2 r_1)}{u(r_2)u(r_1)}$. This quotient systems, which is equal to standard $\omega'_{(1)}(r_2,r_1)$ of $K_1$ class for group 422, is presented in tab.3

**Table 3. Standard quotient system of $K_1$ class for group 422**

| $\omega'_{(1)}(r_2,r_1)$ $[\omega'_1(r_2,r_1), \omega'_2(r_2,r_1)]$ $r_2$ | $r_1$ | $e$ | $c_2$ | $c_4$ | $c_4^3$ | $(u_2)_1$ | $(u_2)_2$ | $(u'_2)_1$ | $(u'_2)_2$ |
|---|---|---|---|---|---|---|---|---|---|
| $e$ |  | 1 | 1 | 1 | 1 | 1 | 1 | 1 | 1 |
| $c_2$ |  | 1 | -1 | 1 | -1 | -1 | 1 | -1 | 1 |
| $c_4$ |  | 1 | 1 | 1 | -1 | 1 | -1 | -1 | -1 |
| $c_4^3$ |  | 1 | -1 | -1 | -1 | 1 | 1 | 1 | -1 |
| $(u_2)_1$ |  | 1 | 1 | 1 | 1 | 1 | 1 | 1 | 1 |
| $(u_2)_2$ |  | 1 | -1 | 1 | -1 | -1 | 1 | -1 | 1 |
| $(u'_2)_1$ |  | 1 | 1 | 1 | -1 | 1 | -1 | -1 | -1 |
| $(u'_2)_2$ |  | 1 | -1 | -1 | -1 | 1 | 1 | 1 | -1 |

The quotient system $\omega_2(r_2,r_1)$ is defined as following:

$$\omega_2(r_2,r_1) = \begin{cases} 1 & \text{if } 0 \leq \theta < 2\pi \\ -1 & \text{if } 2\pi \leq \theta < 4\pi \end{cases}, \quad (6)$$

where $\theta$ is the rotation angle, which corresponds to product of elements $r_2 r_1$. This quotient system is the same for points Γ and Z, because it is determined only by symmetry group of wave vectors directions. Quotient system $\omega_2(r_2,r_1)$ is a conversion factor into spin space. Let us set the group by basis elements and defining relations for obtaining $\omega_2(r_2,r_1)$. Basis elements: $a = c_4$, $b = (u_2)_1$, and defining relations: $a^8 = e$, $b^4 = e$, $ab = qba^3$ (relations like for double group), where $q$ is $2\pi$ rotation around any given axis. Element $q$ commutates with all symmetry elements. So, we have the following: $e = b^0 a^0$, $c_2 = b^0 a^2$, $c_4 = b^0 a^1$, $c_4^3 = b^0 a^3$, $(u_2)_1 = b^1 a^0$, $(u_2)_2 = qb^1 a^2$, $(u'_2)_1 = qb^1 a^1$, $(u'_2)_2 = b^1 a^3$. The quotient system $\omega_2(r_2,r_1)$ for group 422 is shown in tab.4

**Table 4.** The quotient system $\omega_2(r_2,r_1)$ for crystal class 422 (I) and functions $u_2(r)$ (II).

| $\omega_2(r_2,r_1)$ \ $r_1$ / $r_2$ | $e$ | $c_2$ | $c_4$ | $c_4^3$ | $(u_2)_1$ | $(u_2)_2$ | $(u_2')_1$ | $(u_2')_2$ | |
|---|---|---|---|---|---|---|---|---|---|
| $e$ | 1 | 1 | 1 | 1 | 1 | 1 | 1 | 1 | |
| $c_2$ | 1 | -1 | 1 | -1 | 1 | -1 | 1 | -1 | |
| $c_4$ | 1 | 1 | 1 | -1 | -1 | 1 | -1 | -1 | I |
| $c_4^3$ | 1 | -1 | -1 | -1 | 1 | 1 | -1 | 1 | |
| $(u_2)_1$ | 1 | -1 | -1 | 1 | -1 | 1 | 1 | -1 | |
| $(u_2)_2$ | 1 | 1 | -1 | -1 | -1 | -1 | 1 | 1 | |
| $(u_2')_1$ | 1 | -1 | 1 | 1 | -1 | -1 | -1 | 1 | |
| $(u_2')_2$ | 1 | 1 | -1 | 1 | 1 | -1 | -1 | -1 | |
| $u_2(r)$ | 1 | -1 | -i | i | i | i | -1 | -1 | II |

Procedure of reducing the quotient system $\omega_2(r_2,r_1)$ to standard view $\omega_2'(r_2,r_1)$ is similar to procedure of reducing the quotient system $\omega_1(r_2,r_1)$. The quotient system $\omega_2'(r_2,r_1)$ belongs to $K_1$ class as the quotient system $\omega_1'(r_2,r_1)$. This standard quotient system $\omega_{(1)}'(r_2,r_1)$ has already shown in tab.3.

The characters of irreducible projective representations, which correspond to standard quotient system of $K_1$ class [1] are presented in tab.5. Characters of irreducible projective spinor representations (tab.6) can be obtained by multiplication of characters of tab.5 by corresponding functions $u_2(r)$ (these characters are equal to characters of double group (422)').

**Table 5.** Characters of irreducible projective representations, which correspond to standard quotient system of $K_1$ class for group 422

| 422 | $e$ | $c_2$ | $c_4$ | $c_4^3$ | $(u_2)_1$ | $(u_2)_2$ | $(u_2')_1$ | $(u_2')_2$ |
|---|---|---|---|---|---|---|---|---|
| $P_1^{(1)}$ | 2 | 0 | $\sqrt{2}i$ | $\sqrt{2}i$ | 0 | 0 | 0 | 0 |
| $P_2^{(1)}$ | 2 | 0 | $-\sqrt{2}i$ | $-\sqrt{2}i$ | 0 | 0 | 0 | 0 |

**Table 6.** Characters of irreducible projective spinor representations for group 422

| (422)' | | $e$ | $q$ | $c_2$ / $qc_2$ | $c_4$ / $qc_4^3$ | $c_4^3$ / $qc_4$ | $2u_2$ / $2qu_2$ | $2u_2'$ / $2qu_2'$ |
|---|---|---|---|---|---|---|---|---|
| $\Gamma_1$ | $A_1$ | 1 | 1 | 1 | 1 | 1 | 1 | 1 |
| $\Gamma_2$ | $A_2$ | 1 | 1 | 1 | 1 | 1 | -1 | -1 |
| $\Gamma_3$ | $B_1$ | 1 | 1 | 1 | -1 | -1 | 1 | -1 |
| $\Gamma_4$ | $B_2$ | 1 | 1 | 1 | -1 | -1 | -1 | 1 |
| $\Gamma_5$ | $E$ | 2 | 2 | -2 | 0 | 0 | 0 | 0 |
| $\Gamma_6$ | $E_1'$ | 2 | -2 | 0 | $\sqrt{2}$ | $-\sqrt{2}$ | 0 | 0 |
| $\Gamma_7$ | $E_2'$ | 2 | -2 | 0 | $-\sqrt{2}$ | $\sqrt{2}$ | 0 | 0 |

It is necessary to calculate values of $e^{-i\vec{k}(\vec{a}+\vec{a})}$ to construct characters of irreducible projective representations for points $\Gamma$ and Z. Values of vectors $\vec{a}$ (tab.7) for elements of groups P4$_1$2$_1$2 and P4$_3$2$_1$2 can be calculated using above mentioned vectors of non-trivial translations.

**Table 7.** Values of vectors $\vec{a}$ for elements of groups 4$_1$2$_1$2 and 4$_3$2$_1$2.

| Група | $e$ | $c_2$ | $c_4$ | $c_4^3$ | $(u_2)_1$ | $(u_2)_2$ | $(u_2')_1$ | $(u_2')_2$ |
|---|---|---|---|---|---|---|---|---|
| P4$_1$2$_1$2 | 0 | $-\vec{a}_3$ | 0 | $\vec{a}_2$ | 0 | $\vec{a}_3$ | $-\vec{a}_2$ | 0 |
| P4$_3$2$_1$2 | 0 | $-\vec{a}_1-\vec{a}_3$ | 0 | $-2\vec{a}_1+\vec{a}_2$ | 0 | $2\vec{a}_1+\vec{a}_3$ | $\vec{a}_1-\vec{a}_2$ | $2\vec{a}_1$ |

We calculate characters of irreducible projective representations for points $\Gamma$ and Z (tab.8 and 9) by substitution matrix of irreducible projective representations $D(r)$ on their characters $\chi_D(r)$, which correspond to quotient systems of $K_0$ and $K_1$ classes for states with integer and half-integer spin. It should be noted, that characters of irreducible projective representations for space groups, which describe two enantiomorphous modifications are the same.

**Table 8. Characters of irreducible projective representations of wave vector group of group 422 for point Γ**

| 422 | Projective class | $e$ | $c_2$ | $c_4$ | $c_4^3$ | $(u_2)_1$ | $(u_2)_2$ | $(u_2')_1$ | $(u_2')_2$ |
|---|---|---|---|---|---|---|---|---|---|
| $\Gamma_1$ |  | 1 | 1 | 1 | 1 | 1 | 1 | 1 | 1 |
| $\Gamma_2$ |  | 1 | 1 | 1 | 1 | -1 | -1 | -1 | -1 |
| $\Gamma_3$ | $K_0$ | 1 | 1 | -1 | -1 | 1 | 1 | -1 | -1 |
| $\Gamma_4$ |  | 1 | 1 | -1 | -1 | -1 | -1 | 1 | 1 |
| $\Gamma_5$ |  | 2 | -2 | 0 | 0 | 0 | 0 | 0 | 0 |
| $\Gamma_6$ | $K_1$ | 2 | 0 | $\sqrt{2}$ | $-\sqrt{2}$ | 0 | 0 | 0 | 0 |
| $\Gamma_7$ |  | 2 | 0 | $-\sqrt{2}$ | $\sqrt{2}$ | 0 | 0 | 0 | 0 |

**Table 9. Characters of irreducible projective representations of wave vector group of groups P4₁2₁2 and P4₃2₁2 for point Z**

| 422 | Projective class | $e$ | $c_2$ | $c_4$ | $c_4^3$ | $(u_2)_1$ | $(u_2)_2$ | $(u_2')_1$ | $(u_2')_2$ |
|---|---|---|---|---|---|---|---|---|---|
| $Z_1$ |  | 1 | 1 | -1 | -1 | $i$ | $-i$ | $i$ | $-i$ |
| $Z_2$ |  | 1 | 1 | -1 | -1 | $-i$ | $i$ | $-i$ | $i$ |
| $Z_3$ | $K_0$ | 1 | 1 | 1 | 1 | $i$ | $-i$ | $-i$ | $i$ |
| $Z_4$ |  | 1 | 1 | 1 | 1 | $-i$ | $i$ | $i$ | $-i$ |
| $Z_5$ |  | 2 | -2 | 0 | 0 | 0 | 0 | 0 | 0 |
| $Z_6$ | $K_1$ | 2 | 0 | $\sqrt{2}$ | $-\sqrt{2}$ | 0 | 0 | 0 | 0 |
| $Z_7$ |  | 2 | 0 | $-\sqrt{2}$ | $\sqrt{2}$ | 0 | 0 | 0 | 0 |

or

| 422 | Projective class | $e$ | $c_2$ | $c_4$ | $c_4^3$ | $(u_2)_1$ | $(u_2)_2$ | $(u_2')_1$ | $(u_2')_2$ |
|---|---|---|---|---|---|---|---|---|---|
| $(Z_1 + Z_2)$ |  | 2 | 2 | -2 | -2 | 0 | 0 | 0 | 0 |
| $(Z_3 + Z_4)$ | $K_0$ | 2 | 2 | 2 | 2 | 0 | 0 | 0 | 0 |
| $Z_5$ |  | 2 | -2 | 0 | 0 | 0 | 0 | 0 | 0 |
| $Z_6$ | $K_1$ | 2 | 0 | $\sqrt{2}$ | $-\sqrt{2}$ | 0 | 0 | 0 | 0 |
| $Z_7$ |  | 2 | 0 | $-\sqrt{2}$ | $\sqrt{2}$ | 0 | 0 | 0 | 0 |

For points S, A, Σ, M, V, R, and X characters of irreducible projective representations is shown in appendix. The merging and doubling of representations are caused by degeneracy of states due to invariance with respect to time inversion. Ganging and doubling were made according to the Herring criterion [1].

There is a principal possibility to examine transformation characters of irreducible projective representations along $Oz$ axis from point Γ up to point Z. Therefore we have to determine characters of irreducible projective representations for point Λ as a function of wave vector $\vec{k}_\Lambda$ and $0 < \vec{k}_\Lambda < -\frac{\vec{b}_1}{2}$ (tab.10 and 11). As it was mentioned the wave vector group of point Λ is isomorphous to point group 4. Group 4 is sub-group of limiting symmetry group (or Curie group). The limiting symmetry groups include the infinite-fold axis as symmetry element, and these groups are enantiomorphous, subordinate point groups are enantiomorphous too [2].

So, it is necessary to examine right point group 4 (it has screw tetrad axis 4₁) and left one $\widetilde{4}$ (it has screw tetrad axis 4₃). Group 4 has basis element $c_4$ (rotation on angle $\varphi = \frac{2\pi}{4}$ in right-hand coordinate), and group $\widetilde{4}$ – $\widetilde{c}_4 = c_4^3$ (rotation on angle $\varphi = \frac{3\pi}{2}$ in right-hand coordinate). These functions are called correlations of compatibility. They can be used for construction of principal view of dispersion curves.

**Table 10.** Characters of irreducible projective representations of wave vector group of group P4$_1$2$_1$2 for point Λ ($\eta_{\vec{k}} = e^{-i\vec{k}\vec{a}_1/2}$)

| 4 | $e$ | $c_4$ | $c_2$ | $c_4^3$ |
|---|---|---|---|---|
| Λ$_1$ | 1 | $\eta_{\vec{k}}$ | $\eta_{\vec{k}}^2$ | $\eta_{\vec{k}}^3$ |
| Λ$_2$ | 1 | $-\eta_{\vec{k}}$ | $\eta_{\vec{k}}^2$ | $-\eta_{\vec{k}}^3$ |
| Λ$_3$ | 1 | $i\eta_{\vec{k}}$ | $-\eta_{\vec{k}}^2$ | $-i\eta_{\vec{k}}^3$ |
| Λ$_4$ | 1 | $-i\eta_{\vec{k}}$ | $-\eta_{\vec{k}}^2$ | $i\eta_{\vec{k}}^3$ |
| Λ$_5$ | 1 | $\varepsilon_8 \eta_{\vec{k}}$ | $i\eta_{\vec{k}}^2$ | $-\varepsilon_8^{-1}\eta_{\vec{k}}^3$ |
| Λ$_6$ | 1 | $-\varepsilon_8 \eta_{\vec{k}}$ | $i\eta_{\vec{k}}^2$ | $\varepsilon_8^{-1}\eta_{\vec{k}}^3$ |
| Λ$_7$ | 1 | $-\varepsilon_8^{-1}\eta_{\vec{k}}$ | $-i\eta_{\vec{k}}^2$ | $\varepsilon_8 \eta_{\vec{k}}^3$ |
| Λ$_8$ | 1 | $\varepsilon_8^{-1}\eta_{\vec{k}}$ | $-i\eta_{\vec{k}}^2$ | $-\varepsilon_8 \eta_{\vec{k}}^3$ |

**Table 11.** Characters of irreducible projective representations of wave vector group of group P4$_3$2$_1$2 for point Λ ($\eta_{\vec{k}} = e^{-i\vec{k}\vec{a}_1/2}$)

| $\tilde{4}$ | $\tilde{e}$ | $\tilde{c}_4$ | $\tilde{c}_2$ | $\tilde{c}_4^3$ |
|---|---|---|---|---|
| $\tilde{\Lambda}_1$ | 1 | $\eta_{\vec{k}}$ | $\eta_{\vec{k}}^2$ | $\eta_{\vec{k}}^3$ |
| $\tilde{\Lambda}_2$ | 1 | $-\eta_{\vec{k}}$ | $\eta_{\vec{k}}^2$ | $-\eta_{\vec{k}}^3$ |
| $\tilde{\Lambda}_3$ | 1 | $-i\eta_{\vec{k}}$ | $-\eta_{\vec{k}}^2$ | $i\eta_{\vec{k}}^3$ |
| $\tilde{\Lambda}_4$ | 1 | $i\eta_{\vec{k}}$ | $-\eta_{\vec{k}}^2$ | $-i\eta_{\vec{k}}^3$ |
| $\tilde{\Lambda}_5$ | 1 | $-\varepsilon_8^{-1}\eta_{\vec{k}}$ | $-i\eta_{\vec{k}}^2$ | $\varepsilon_8 \eta_{\vec{k}}^3$ |
| $\tilde{\Lambda}_6$ | 1 | $\varepsilon_8^{-1}\eta_{\vec{k}}$ | $-i\eta_{\vec{k}}^2$ | $-\varepsilon_8 \eta_{\vec{k}}^3$ |
| $\tilde{\Lambda}_7$ | 1 | $\varepsilon_8 \eta_{\vec{k}}$ | $i\eta_{\vec{k}}^2$ | $-\varepsilon_8^{-1}\eta_{\vec{k}}^3$ |
| $\tilde{\Lambda}_8$ | 1 | $-\varepsilon_8 \eta_{\vec{k}}$ | $i\eta_{\vec{k}}^2$ | $\varepsilon_8^{-1}\eta_{\vec{k}}^3$ |

Thus, the method, we suggest, gives the possibility to examine any point of Brilluin zone, to find characters of irreducible projective representations, and to determinate functions of wave vector, which describe transformations representations from one point to another. All these data can be used for determination of selections rules of different types of energy excitations, for analysing whether examined points are points of zero slope and for construction of dispersion curves for different directions of wave vector.

**APPENDIX**

**Table 1.** Characters of irreducible projective representations of wave vector group of groups P4$_1$2$_1$2 and P4$_3$2$_1$2 for point A

| 422 ($D_4$) | Projective class | $e$ | $c_2$ | $c_4$ | $c_4^3$ | $(u_2)_1$ | $(u_2)_2$ | $(u_2')_1$ | $(u_2')_2$ |
|---|---|---|---|---|---|---|---|---|---|
| A$_1$ |  | 1 | -1 | $-i$ | $i$ | 1 | 1 | $i$ | $i$ |
| A$_2$ |  | 1 | -1 | $-i$ | $i$ | -1 | -1 | $-i$ | $-i$ |
| A$_3$ | K$_0$ | 1 | -1 | $i$ | $-i$ | 1 | 1 | $-i$ | $-i$ |
| A$_4$ |  | 1 | -1 | $i$ | $-i$ | -1 | -1 | $i$ | $i$ |
| A$_5$ |  | 2 | 2 | 0 | 0 | 0 | 0 | 0 | 0 |
| A$_6$ | K$_1$ | 2 | 0 | $\sqrt{2}i$ | $\sqrt{2}i$ | 0 | 0 | 0 | 0 |
| A$_7$ |  | 2 | 0 | $-\sqrt{2}i$ | $-\sqrt{2}i$ | 0 | 0 | 0 | 0 |

or

| | | | | | | | | | |
|---|---|---|---|---|---|---|---|---|---|
| (A$_1$ + A$_3$) |  | 2 | -2 | 0 | 0 | 2 | 2 | 0 | 0 |
| (A$_2$ + A$_4$) | K$_0$ | 2 | -2 | 0 | 0 | -2 | -2 | 0 | 0 |
| ((A$_5$)) |  | 4 | 4 | 0 | 0 | 0 | 0 | 0 | 0 |
| (A$_6$ + A$_7$) | K$_1$ | 4 | 0 | 0 | 0 | 0 | 0 | 0 | 0 |

**Table 2.** Characters of irreducible projective representations of wave vector group of groups P4$_1$2$_1$2 and P4$_3$2$_1$2 for point M

| P4$_1$2$_1$2 та P4$_3$2$_1$2 | Projective class | $e$ | $c_2$ | $c_4$ | $c_4^3$ | $(u_2)_1$ | $(u_2)_2$ | $(u_2')_1$ | $(u_2')_2$ |
|---|---|---|---|---|---|---|---|---|---|
| M$_1$ |  | 1 | -1 | $i$ | $-i$ | $i$ | $-i$ | -1 | 1 |
| M$_2$ |  | 1 | -1 | $i$ | $-i$ | $-i$ | $i$ | 1 | -1 |
| M$_3$ | K$_0$ | 1 | -1 | $-i$ | $i$ | $i$ | $-i$ | 1 | -1 |
| M$_4$ |  | 1 | -1 | $-i$ | $i$ | $-i$ | $i$ | -1 | 1 |
| M$_5$ |  | 2 | 2 | 0 | 0 | 0 | 0 | 0 | 0 |
| M$_6$ | K$_1$ | 2 | 0 | $i\sqrt{2}$ | $i\sqrt{2}$ | 0 | 0 | 0 | 0 |
| M$_7$ |  | 2 | 0 | $-i\sqrt{2}$ | $-i\sqrt{2}$ | 0 | 0 | 0 | 0 |

or

| | | | | | | | | | |
|---|---|---|---|---|---|---|---|---|---|
| (M$_1$ + M$_4$) |  | 2 | -2 | 0 | 0 | 0 | 0 | -2 | 2 |
| (M$_2$ + M$_3$) | K$_0$ | 2 | -2 | 0 | 0 | 0 | 0 | 2 | -2 |
| M$_5$ |  | 2 | 2 | 0 | 0 | 0 | 0 | 0 | 0 |
| (M$_6$ + M$_7$) | K$_1$ | 4 | 0 | 0 | 0 | 0 | 0 | 0 | 0 |

**Table 3.** Characters of irreducible projective representations of wave vector group of groups P4$_1$2$_1$2 and P4$_3$2$_1$2 for point X

| P4$_1$2$_1$2 та P4$_3$2$_1$2 | Projective class | $e$ | $c_2$ | $(u_2)_1$ | $(u_2)_2$ |
|---|---|---|---|---|---|
| X$_1$ |  | 1 | $-i$ | 1 | $i$ |
| X$_2$ | K$_0$ | 1 | $-i$ | -1 | $-i$ |
| X$_3$ |  | 1 | $i$ | 1 | $-i$ |
| X$_4$ |  | 1 | $i$ | -1 | $i$ |
| X$_5$ | K$_1$ | 2 | 0 | 0 | 0 |

or

| | | | | | |
|---|---|---|---|---|---|
| (X$_1$+X$_3$) | K$_0$ | 2 | 0 | 2 | 0 |
| (X$_2$+X$_4$) |  | 2 | 0 | -2 | 0 |
| X$_5$ | K$_1$ | 2 | 0 | 0 | 0 |

**Table 4.** Characters of irreducible projective representations of wave vector group of groups $P4_12_12$ and $P4_32_12$ for point R

| $P4_12_12$ | Projective class | $e$ | $c_2$ | $(u_2)_1$ | $(u_2)_2$ |
|---|---|---|---|---|---|
| $R_1$ | | 1 | $-i$ | $-i$ | 1 |
| $R_2$ | $K_0$ | 1 | $-i$ | $-i$ | -1 |
| $R_3$ | | 1 | $i$ | $i$ | -1 |
| $R_4$ | | 1 | $i$ | $i$ | 1 |
| $R_5$ | $K_1$ | 2 | 0 | 0 | 0 |

or

| | | | | | |
|---|---|---|---|---|---|
| $(R_1+R_4)$ | $K_0$ | 2 | 0 | 0 | 2 |
| $(R_2+R_3)$ | | 2 | 0 | 0 | -2 |
| $((R_5))$ | $K_1$ | 2 | 0 | 0 | 0 |

**Table 5.** Characters of irreducible projective representations of wave vector group of group $P4_12_12$ for point V ($\eta_{\vec{k}} = e^{-i\vec{k}\vec{a}_1/4}$, $\vec{k} \in \left(-\frac{b_2}{2} - \frac{b_3}{2}, -\frac{b_1}{2} - \frac{b_2}{2} - \frac{b_3}{2}\right)$)

| $4(C_4)$ | $e$ | $c_4$ | $c_2$ | $c_4^3$ |
|---|---|---|---|---|
| $V_1$ | 1 | $i\eta_{\vec{k}}$ | $-\eta_{\vec{k}}^2$ | $-i\eta_{\vec{k}}^3$ |
| $V_2$ | 1 | $-i\eta_{\vec{k}}$ | $-\eta_{\vec{k}}^2$ | $i\eta_{\vec{k}}^3$ |
| $V_3$ | 1 | $-\eta_{\vec{k}}$ | $\eta_{\vec{k}}^2$ | $-\eta_{\vec{k}}^3$ |
| $V_4$ | 1 | $\eta_{\vec{k}}$ | $\eta_{\vec{k}}^2$ | $\eta_{\vec{k}}^3$ |
| $V_5$ | 1 | $-\varepsilon_8^{-1}\eta_{\vec{k}}$ | $-i\eta_{\vec{k}}^2$ | $\varepsilon_8\eta_{\vec{k}}^3$ |
| $V_6$ | 1 | $\varepsilon_8^{-1}\eta_{\vec{k}}$ | $-i\eta_{\vec{k}}^2$ | $-\varepsilon_8\eta_{\vec{k}}^3$ |
| $V_7$ | 1 | $-\varepsilon_8\eta_{\vec{k}}$ | $i\eta_{\vec{k}}^2$ | $\varepsilon_8^{-1}\eta_{\vec{k}}^3$ |
| $V_8$ | 1 | $\varepsilon_8\eta_{\vec{k}}$ | $i\eta_{\vec{k}}^2$ | $-\varepsilon_8^{-1}\eta_{\vec{k}}^3$ |

**Table 6.** Characters of irreducible projective representations of wave vector group of group $P4_12_12$ for point V ($\eta_{\vec{k}} = e^{-i\vec{k}\vec{a}_1/4}$, $\vec{k} \in \left(-\frac{b_2}{2} - \frac{b_3}{2}, -\frac{b_1}{2} - \frac{b_2}{2} - \frac{b_3}{2}\right)$)

| $\widetilde{4}(\widetilde{C}_4)$ | $\widetilde{e}$ | $\widetilde{c}_4$ | $\widetilde{c}_2$ | $\widetilde{c}_4^3$ |
|---|---|---|---|---|
| $\widetilde{V}_1$ | 1 | $-i\eta_{\vec{k}}$ | $-\eta_{\vec{k}}^2$ | $i\eta_{\vec{k}}^3$ |
| $\widetilde{V}_2$ | 1 | $i\eta_{\vec{k}}$ | $-\eta_{\vec{k}}^2$ | $-i\eta_{\vec{k}}^3$ |
| $\widetilde{V}_3$ | 1 | $-\eta_{\vec{k}}$ | $\eta_{\vec{k}}^2$ | $-\eta_{\vec{k}}^3$ |
| $\widetilde{V}_4$ | 1 | $\eta_{\vec{k}}$ | $\eta_{\vec{k}}^2$ | $\eta_{\vec{k}}^3$ |
| $\widetilde{V}_5$ | 1 | $\varepsilon_8\eta_{\vec{k}}$ | $-i\eta_{\vec{k}}^2$ | $-\varepsilon_8^{-1}\eta_{\vec{k}}^3$ |
| $\widetilde{V}_6$ | 1 | $-\varepsilon_8\eta_{\vec{k}}$ | $-i\eta_{\vec{k}}^2$ | $\varepsilon_8^{-1}\eta_{\vec{k}}^3$ |
| $\widetilde{V}_7$ | 1 | $\varepsilon_8^{-1}\eta_{\vec{k}}$ | $i\eta_{\vec{k}}^2$ | $-\varepsilon_8\eta_{\vec{k}}^3$ |
| $\widetilde{V}_8$ | 1 | $-\varepsilon_8^{-1}\eta_{\vec{k}}$ | $i\eta_{\vec{k}}^2$ | $\varepsilon_8\eta_{\vec{k}}^3$ |

**Table 7.** Characters of irreducible projective representations of wave vector group of groups $P4_12_12$ and $P4_32_12$ for point Σ ($\eta_{\vec{k}} = e^{-i\vec{k}\vec{a}_2}$)

| $\vec{k} \in \left(0, -\frac{\vec{b}_2}{2} - \frac{\vec{b}_3}{2}\right)$ | | | | $\vec{k} \in \left(0, \frac{\vec{b}_2}{2} - \frac{\vec{b}_3}{2}\right)$ | | |
|---|---|---|---|---|---|---|
| $2(C_2)$ | $e$ | $(u'_2)_2$ | | $2(C_2)$ | $e$ | $(u'_2)_2$ |
| $\Sigma_1$ | 1 | 1 | | $\Sigma_1$ | 1 | 1 |
| $\Sigma_2$ | 1 | -1 | | $\Sigma_2$ | 1 | -1 |
| $\Sigma_3$ | 1 | $i$ | | $\Sigma_3$ | 1 | $i$ |
| $\Sigma_4$ | 1 | $-i$ | | $\Sigma_4$ | 1 | $-i$ |

| $\vec{k} \in \left(0, \frac{\vec{b}_2}{2} + \frac{\vec{b}_3}{2}\right)$ | | | | $\vec{k} \in \left(0, -\frac{\vec{b}_2}{2} + \frac{\vec{b}_3}{2}\right)$ | | |
|---|---|---|---|---|---|---|
| $2(C_2)$ | $e$ | $(u'_2)_1$ | | $2(C_2)$ | $e$ | $(u'_2)_1$ |
| $\Sigma_1$ | 1 | $\eta^*_{\vec{k}}$ | | $\Sigma_1$ | 1 | $\eta_{\vec{k}}$ |
| $\Sigma_2$ | 1 | $-\eta^*_{\vec{k}}$ | | $\Sigma_2$ | 1 | $-\eta_{\vec{k}}$ |
| $\Sigma_3$ | 1 | $i\eta^*_{\vec{k}}$ | | $\Sigma_3$ | 1 | $i\eta_{\vec{k}}$ |
| $\Sigma_4$ | 1 | $-i\eta^*_{\vec{k}}$ | | $\Sigma_4$ | 1 | $-i\eta_{\vec{k}}$ |

**Table 8.** Characters of irreducible projective representations of wave vector group of groups $P4_12_12$ for point S ($\eta_{\vec{k}} = e^{-i\vec{k}\vec{a}_2}$)

| $\vec{k} \in \left(-\frac{\vec{b}_1}{2}, -\frac{\vec{b}_1}{2} - \frac{\vec{b}_2}{2} - \frac{\vec{b}_3}{2}\right)$ | | | | $\vec{k} \in \left(-\frac{\vec{b}_1}{2}, -\frac{\vec{b}_1}{2} + \frac{\vec{b}_2}{2} - \frac{\vec{b}_3}{2}\right)$ | | |
|---|---|---|---|---|---|---|
| $2(C_2)$ | $e$ | $(u'_2)_2$ | | $2(C_2)$ | $e$ | $(u'_2)_2$ |
| $S_1$ | 1 | 1 | | $S_1$ | 1 | 1 |
| $S_2$ | 1 | -1 | | $S_2$ | 1 | -1 |
| $S_3$ | 1 | $i$ | | $S_3$ | 1 | $i$ |
| $S_4$ | 1 | $-i$ | | $S_4$ | 1 | $-i$ |

| $\vec{k} \in \left(-\frac{\vec{b}_1}{2}, -\frac{\vec{b}_1}{2} + \frac{\vec{b}_2}{2} + \frac{\vec{b}_3}{2}\right)$ | | | | $\vec{k} \in \left(-\frac{\vec{b}_1}{2}, -\frac{\vec{b}_1}{2} - \frac{\vec{b}_2}{2} + \frac{\vec{b}_3}{2}\right)$ | | |
|---|---|---|---|---|---|---|
| $2(C_2)$ | $e$ | $(u'_2)_1$ | | $2(C_2)$ | $e$ | $(u'_2)_1$ |
| $S_1$ | 1 | $\eta^*_{\vec{k}}$ | | $S_1$ | 1 | $\eta_{\vec{k}}$ |
| $S_2$ | 1 | $-\eta^*_{\vec{k}}$ | | $S_2$ | 1 | $-\eta_{\vec{k}}$ |
| $S_3$ | 1 | $i\eta^*_{\vec{k}}$ | | $S_3$ | 1 | $i\eta_{\vec{k}}$ |
| $S_4$ | 1 | $-i\eta^*_{\vec{k}}$ | | $S_4$ | 1 | $-i\eta_{\vec{k}}$ |

**Table 8.** Characters of irreducible projective representations of wave vector group of groups $P4_32_12$ for point S ($\eta_{\vec{k}} = e^{-i\vec{k}\vec{a}_2}$)

| $\vec{k} \in \left(-\frac{\vec{b}_1}{2}, -\frac{\vec{b}_1}{2} - \frac{\vec{b}_2}{2} - \frac{\vec{b}_3}{2}\right)$ | | | $\vec{k} \in \left(-\frac{\vec{b}_1}{2}, -\frac{\vec{b}_1}{2} + \frac{\vec{b}_2}{2} - \frac{\vec{b}_3}{2}\right)$ | | |
|---|---|---|---|---|---|
| $2(C_2)$ | $e$ | $(u'_2)_2$ | $2(C_2)$ | $e$ | $(u'_2)_2$ |
| $S_1$ | 1 | 1 | $S_1$ | 1 | 1 |
| $S_2$ | 1 | -1 | $S_2$ | 1 | -1 |
| $S_3$ | 1 | $i$ | $S_3$ | 1 | $i$ |
| $S_4$ | 1 | $-i$ | $S_4$ | 1 | $-i$ |
| $\vec{k} \in \left(-\frac{\vec{b}_1}{2}, -\frac{\vec{b}_1}{2} + \frac{\vec{b}_2}{2} + \frac{\vec{b}_3}{2}\right)$ | | | $\vec{k} \in \left(-\frac{\vec{b}_1}{2}, -\frac{\vec{b}_1}{2} - \frac{\vec{b}_2}{2} + \frac{\vec{b}_3}{2}\right)$ | | |
| $2(C_2)$ | $e$ | $(u'_2)_1$ | $2(C_2)$ | $e$ | $(u'_2)_1$ |
| $S_1$ | 1 | $\eta^*_{\vec{k}}$ | $S_1$ | 1 | $\eta_{\vec{k}}$ |
| $S_2$ | 1 | $-\eta^*_{\vec{k}}$ | $S_2$ | 1 | $-\eta_{\vec{k}}$ |
| $S_3$ | 1 | $-i\eta^*_{\vec{k}}$ | $S_3$ | 1 | $-i\eta_{\vec{k}}$ |
| $S_4$ | 1 | $i\eta^*_{\vec{k}}$ | $S_4$ | 1 | $i\eta_{\vec{k}}$ |